\documentclass[twocolumn,preprintnumbers,amsmath,amssymb]{revtex4}
\usepackage{graphicx}
\usepackage{dcolumn}
\usepackage{bm}

\newcommand{\comment}[1]{}
\newcommand{\li}{$^{6}$Li }
\newcommand{\yb}{$^{174}$Yb }

\usepackage{color}

\begin{document}

\title{Sympathetic cooling in an optically trapped mixture of alkali and spin-singlet atoms}

\author{Vladyslav V. Ivanov}
\author{Alexander Khramov}
\author{Anders H. Hansen}
\author{William H. Dowd}
\author{Frank M$\ddot{\rm u}$nchow}
\author{Alan O. Jamison}
\author{Subhadeep Gupta}
\affiliation{Department of Physics, University of Washington, Seattle WA 98195}

\date{\today}

\begin{abstract}

We report on the realization of a stable mixture of ultracold lithium and ytterbium atoms confined in a far-off-resonance 
optical dipole trap. We observe sympathetic cooling of \li by \yb and extract the $s$-wave scattering length magnitude 
$|a_{^6{\rm Li}-^{174}{\rm Yb}}|=(13 \pm 3)a_0$ from the rate of inter-species thermalization. Using forced 
evaporative cooling of \yb\!, we achieve reduction of the \li temperature to below the Fermi temperature, purely through 
inter-species sympathetic cooling.

\end{abstract}

\maketitle

Ultracold mixtures composed of different atomic species \cite{modu01,hadz02,silb05,aubi06,tagl08,spie09} offer unique 
opportunities for probing few- and many-body physics. These include studies of Efimov states with mass-mismatched 
collision partners \cite{dinc06,marc08}, impurity probes of superfluid properties \cite{spie09,vern10}, and mass 
imbalanced regimes of interactions and pairing in Fermi gases \cite{iski08,geze09,tren10}. Further, the components of the 
mixture can be linked through field-induced scattering resonances to produce heteronuclear molecules 
\cite{sage05,dieg08,ni08}, which are expected to be valuable tools for the study of dipolar quantum matter, quantum 
information science, and tests of fundamental physics \cite{carr09}. An essential requirement
for all of these ultracold mixture studies is an understanding of the ground state scattering properties. Favorable 
collisional properties are needed for mixture production and stability, while knowledge of the underlying 
potentials allow identification of regimes of tunable interactions.

In this paper, we report on successful simultaneous optical trapping and measurements of scattering properties for a 
mixture of alkali \li and spin-singlet \yb\!. We observe collisional stability in this mixture and determine the 
magnitude of the previously unknown \li-\yb $s$-wave scattering length from the timescale of inter-species thermalization 
\cite{mudr02,silb05,aubi06}. Furthermore, we sympathetically cool \li to below its Fermi temperature by forced evaporative 
cooling of \yb\!. Unlike the case for bi-alkali mixtures \cite{hadz02,silb05}, our method of
sympathetic cooling an alkali by a spin-singlet atom has the advantage of being immune to inelastic spin changing 
collisions.

While studies of ultracold molecule formation from two-species mixtures are dominated by alkali+alkali 
combinations, molecules created from alkali+spin-singlet mixtures offer the additional advantage of possessing an 
unpaired electron spin to form a paramagnetic ground state. This is a feature of considerable interest for several 
proposed applications including quantum simulations of lattice spin models \cite{mich06}, topological quantum
computing, and sensitive measurements of the electron electric dipole moment \cite{huds02}. Prior to this work a 
dual-species Li-Yb magneto-optical trap (MOT) was demonstrated \cite{okan10}, but with densities too 
low to observe inter-species effects. First results for alkali+spin-singlet mixtures have been 
reported for the Rb+Yb combination. These include photoassociation in a dual-species MOT \cite{nemi09}, and 
observations of sympathetic cooling \cite{tass10} and spatial separation \cite{baum10} in a combined optical and magnetic 
trap.

\begin{figure}
\includegraphics[angle = 0, width = 0.5 \textwidth] {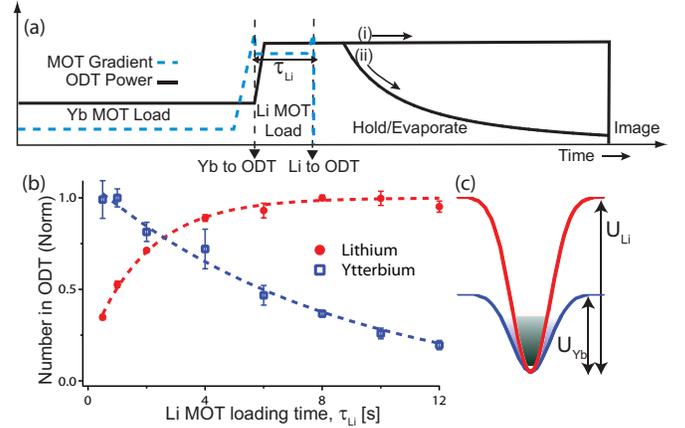}
\vspace{-1cm}
\caption{(color online). Simultaneous optical trapping of \li and \yb\!. (a) shows the typical experimental sequence where 
the laser cooling and ODT loading are performed sequentially for the two species. In addition to the standard procedures 
for single species operation, the magnetic field gradients and the initial optical trap depths are adjusted for each 
species to optimize number and temperature. After both species are in the ODT, either (i) the power is held constant to 
study inter-species thermalization, or (ii) the power is ramped down to perform forced evaporative cooling. Finally, the
optical trap is switched off and the remaining atoms detected with resonant absorption imaging. As displayed in (b), 
varying the Li MOT loading time $\tau_{\text{Li}}$ allows us to control the initial ratio of the two species in the 
optical trap. The dashed lines are exponential fits. The Yb decay time is substantially shorter than the background 
lifetime because of a partial overlap with the \li MOT during $\tau_{\text{Li}}$. (c) Optical trapping potentials for Li 
(red) and Yb (blue) for a given ODT power. The shading depicts the distributions at the same temperature.} 
\label{fig:potsandseqs}
\vspace{-0.5cm}
\end{figure}

In addition to the pursuit of heteronuclear paramagnetic LiYb molecules, our system forms a starting point for studies of 
the \li Fermi superfluid \cite{zwie05} using Yb as an impurity probe. Furthermore, tunable inter-species interactions 
may be induced between Li and Yb through magnetic \cite{zuch10} or optical \cite{ciur05} Feshbach resonances. Together 
with the straightforward availability of fermionic Yb isotopes, this will allow future explorations of few-body collision 
physics in the highly mass-mismatched regime \cite{dinc06,marc08} and fermionic interactions and pairing in mass 
imbalanced mixtures \cite{tren10,geze09}.

Our dual-species experimental setup consists of \yb and \li MOTs which are loaded from separate atomic beams, each 
emerging from a single-species oven and slowed by a single-species Zeeman slower. All laser cooling and absorption 
imaging of \li is performed on the ${^2S_{1/2}} \rightarrow {^2P_{3/2}}$ line (wavelength $\lambda = 671\,$nm, linewidth 
$\Gamma/2\pi = 6\,$MHz). For \yb, we use the ${^1S_0} \rightarrow {^1P_1}$ line ($\lambda = 399\,$nm, 
$\Gamma/2\pi = 29\,$MHz) for Zeeman slowing and absorption imaging, and the ${^1S_0} \rightarrow {^3P_1}$ line 
($\lambda = 556\,$nm, $\Gamma/2\pi = 182\,$kHz) for the MOT. The optical dipole trap (ODT) is derived from the linearly 
polarized output of a 1064nm fiber laser (IPG Photonics YLR-100-1064-LP) and is operated either in single or crossed beam 
geometry. The trap depth is controlled by an acousto-optic modulator.

To mitigate the strong inelastic losses in simultaneous two-species MOTs and to allow for different optimum MOT magnetic 
field gradients \cite{MOTinelnote}, we employ a sequential cooling and trapping strategy (see 
Fig. \ref{fig:potsandseqs}(a)) in which first Yb and then Li is laser cooled and transferred to the ODT. The laser cooling 
sequence for each species consists of a loading phase with large intensities and detunings of the cooling beams, and a 
compression phase where these intensities and detunings are reduced and the MOT gradient increased. The compressed Yb MOT 
contains $\gtrsim2 \times 10^6$ atoms at a temperature of $\lesssim30\,\mu$K. The compressed Li MOT contains 
$\gtrsim10^8$ atoms at $\lesssim400\mu$K, and is optically pumped into the lower $F=1/2$ hyperfine state. Each species is 
transferred to the ODT by using magnetic bias fields to overlap the MOT with the ODT center and then switching off the 
laser cooling beams. 
During the Li laser cooling phase, Yb atoms trapped in the ODT are insensitive to the magnetic field manipulations used 
for the Li MOT. After all cooling beams are switched off, the ODT contains a mixture of Yb and Li atoms 
(see Fig. \ref{fig:potsandseqs}(b)). While all the \yb atoms are in the single 
${^1S_0}$ ground state, the \li atoms are distributed equally between the two $F=1/2$ Zeeman ground states. 

For our ODT wavelength, the relative trap depths and frequencies for the two species are 
$U_{\rm Li}/U_{\rm Yb}=2.2$ and 
$\omega_{\rm Li}/\omega_{\rm Yb}=\sqrt{\frac{(U_{\rm Li}/m_{\rm Li})}{(U_{\rm Yb}/m_{\rm Yb})}}=8$. The relative 
linear size in the harmonic regime is 
$x_{\rm Li}/x_{\rm Yb}=\sqrt{\frac{(T_{\rm Li}/U_{\rm Li})}{(T_{\rm Yb}/U_{\rm Yb})}}=0.7$ for equal 
temperatures (see Fig. \ref{fig:potsandseqs}(c)). The parameters are thus well suited for sympathetic cooling of 
lithium by ytterbium.

To monitor atom number and temperature, we quickly switch off the trap and perform resonant absorption imaging of both 
species for each experimental iteration. Each species is imaged onto a different part of the same CCD camera with 
independently adjustable ballistic expansion times. The trapping potential is characterized through measurements of trap 
frequencies by exciting dipole and breathing oscillations, as well as by parametric heating. The ODT is kept on all the 
time except during imaging.

\begin{figure}
\includegraphics[angle = 0, width = 0.5 \textwidth] {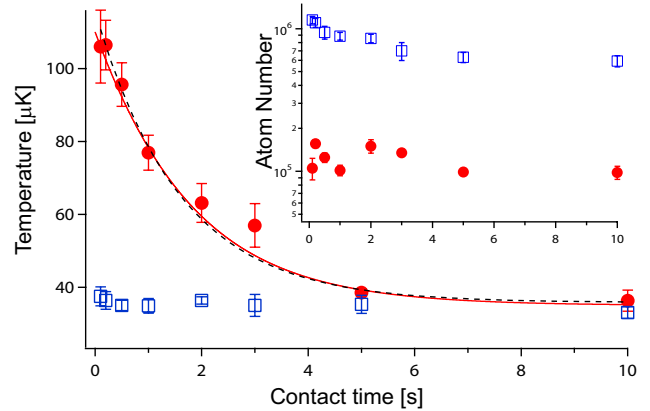}
\vspace{-0.5cm}
\caption{(color online). Sympathetic cooling of \li (red filled circles) by thermalization with a cold \yb bath (blue 
open squares). The temperatures equilibrate with an exponential time constant of $(1.7\pm0.2)\,$s (black dashed line). The 
red solid line is the result of a numerical model (see text). The inset shows the numbers of the two species are almost an 
order of magnitude apart and do not change appreciably during the 2 second thermalization time scale.} 
\label{fig:therm}
\vspace{-0.5cm}
\end{figure}

We first describe our thermalization measurements, which allow us to determine the magnitude of the inter-species $s$-wave 
scattering length. For these measurements, we use a single beam ODT with $1/e^2$ intensity-radius (waist) of 
$30\,\mu$m. Yb is transferred from the MOT into an ODT of calculated depth $U_{\rm Yb} = 220\,\mu$K, after which the 
depth is increased adiabatically in $0.2\,$s to $U_{\rm Yb} = 500\,\mu$K while the Li MOT is loaded 
(see Fig.\ref{fig:potsandseqs}). After $\simeq0.5\,$s MOT loading, Li is transferred into the ODT 
($U_{\rm Li} = 1.1\,$mK). We make our thermalization measurements at this point, where the measured trapping 
frequencies for Yb are $2\pi \times 1600\,$Hz radially and $2\pi \times 13\,$Hz axially. The initial number, 
temperature, and peak density of Yb (Li) atoms are $N_{\rm Yb(Li)}=1.1 \times 10^6 (1.4 \times 10^5)$, 
$T_{\rm Yb(Li)}=35 (110)\,\mu$K and $n_{0,{\rm Yb(Li)}}=1.1 \times 10^{13} (8.4 \times 10^{11})\,$cm$^{-3}$. All the 
thermalization measurements are performed at near-zero magnetic field.

We observe the number and temperature evolution of the two atomic species either in separate single-species experiments, 
or together when in thermal contact with each other. The measured background single-species $1/e$ lifetimes are $>30$ 
seconds for both Li and Yb. The Yb temperature evolution is independent of the presence or absence of Li, equilibrating 
quickly and staying at $35\mu$K throughout the measurement. When Yb is not loaded, the 
two-spin state Li mixture behaves like an ideal gas in the optical trap and remains at its initial temperature of 
$\simeq 100\mu$K without observable changes. This is due to a lack of intrastate collisions from Pauli blocking and a lack
of interstate collisions from the negligible zero-field scattering cross-section \cite{houb98}.
When both species are loaded, the hotter Li cloud equilibrates to the temperature of the Yb cloud (see 
Fig. \ref{fig:therm}). We also observe no change in the Li lifetime from contact with Yb, indicative of negligible 
inelastic interactions between the two species at these experimental parameters. In similar studies at various 
loading parameters, we observe additional losses in the Yb number only. We interpret this to be a result of a
``sympathetic evaporation" effect \cite{mudr02} where during the thermalization process, Li which is confined by a deeper
trap, transfers energy through elastic collisions to Yb and subsequently ejects it from its shallower confinement 
(see Fig. \ref{fig:potsandseqs}(c)). Indeed for larger fractional presence of Li, this effect becomes more pronounced and 
leads to a substantially reduced lifetime for the Yb cloud.

We analyze the thermalization measurements shown in Fig.\ref{fig:therm} by assuming that the elastic interactions are
purely $s$-wave in nature. We are justified in this assumption because our measured temperatures ($\leq 110\mu$K) are
much smaller than the $p-$wave threshold given by $\frac{2}{\sqrt{C_6}} (\frac{\hbar^2 l(l+1)}{6\mu})^{3/2} \simeq 2.5$mK,
where $\mu$ is the reduced mass and using the $C_6$ coefficient for LiYb calculated in \cite{zhan10}. The
thermalization rate $\gamma_{\rm th}$ which characterizes the instantaneous variation of the temperature difference
$\Delta T=T_{\rm Li}-T_{\rm Yb}$, can then be connected to the $s$-wave scattering cross-section $\sigma_{\rm LiYb}$
through the relation:
\begin{equation}
-\frac{1}{\Delta T}\frac{d(\Delta T)}{dt}=\gamma_{\rm th} = \frac{\xi}{\alpha} {\bar n} \sigma_{\rm LiYb} {\bar v}
\label{eqn:collrate}
\end{equation}
Here $\alpha=2.7$ is the average number of collisions needed for thermalization for equal mass partners,
$\xi=\frac{4m_{\rm Li}m_{\rm Yb}}{(m_{\rm Li}+m_{\rm Yb})^2}=0.13$ is the correction factor for non-equal mass collisions 
\cite{mudr02}, ${\bar v}=\sqrt{\frac{8k_B}{\pi}(\frac{T_{\rm Li}}{m_{\rm Li}}+\frac{T_{\rm Yb}}{m_{\rm Yb}})}$ is the 
mean relative velocity, ${\bar n}=(\frac{1}{N_{\rm Li}}+\frac{1}{N_{\rm Yb}})\int n_{\rm Li} n_{\rm Yb} d^3r$ is the 
overlap density, and $n$, $N$, and $T$ are the density, number, and temperature of the two species.
Since $\bar v$ and $\bar n$ change with the Li temperature, we use a numerical procedure to model the thermalization 
process. For a particular value of $s$-wave scattering length $a$, we iterate Eqn.\ref{eqn:collrate} with a short time 
step. We also include an energy dependence \cite{flam99} to the $s$-wave cross-section 
$\sigma_{\rm LiYb}=\frac{4\pi a^2}{(1-(1/2) k^2 r_e a)^2 + k^2a^2}$, where $\hbar k$ is the relative momentum and $r_e$ 
is the effective range evaluated from $C_6$ and $a$. By varying $a$, we obtain a best fit (see 
Fig.\ref{fig:therm}) and infer $|a_{^6{\rm Li}-^{174}{\rm Yb}}|=(13 \pm 3)a_0$. The main limitation to our accuracy is 
the determination of absolute densities. This inaccuracy dominates over variations in $\sigma_{\rm LiYb}$ from the energy 
dependent terms.

Since \yb is spinless, we expect that $a_{^6{\rm Li}-^{174}{\rm Yb}}$ is the same for all hyperfine ground states of 
\li\!. By using state selective imaging at magnetic fields near $500\,$G, we have verified that the two participating \li 
$F=1/2$ Zeeman states maintain equal population and temperature at an intermediate point during sympathetic cooling. We 
would also naively expect that the elastic scattering properties do not change with external 
magnetic field. However a recent theoretical study predicts the existence of magnetically tunable Feshbach resonances in 
alkali+spin-singlet collisions \cite{zuch10}. 

\begin{figure}
\includegraphics[angle = 0, width = 0.5 \textwidth] {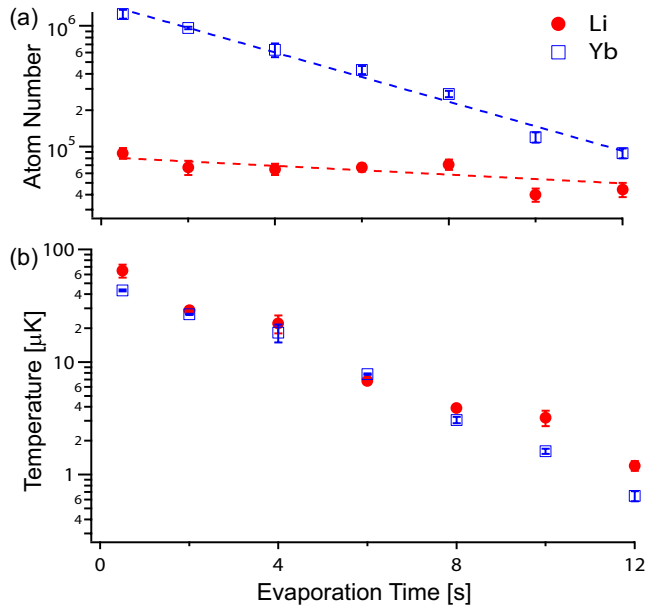}
\vspace{-0.5cm}
\caption{(color online). Sympathetic cooling of lithium by forced evaporative cooling of ytterbium. (a) Number and (b) 
temperature evolution for \li (filled circles) and \yb (open squares) as the power in the crossed optical dipole trap is 
reduced by a factor of 28 over 12 seconds, corresponding to an approximately exponential ramp with a time constant of 
$2.9\,$s.}
\label{fig:symp}
\vspace{-0.5cm}
\end{figure}

Since several potential applications of the LiYb mixture are at conditions near or below quantum degeneracy, we also 
assess methods of increasing the phase space density in the mixture. In single-species \li experiments in our apparatus, 
up to $N_{\rm Li}=2 \times 10^6$ can be loaded into a high power crossed beam ODT. With subsequent forced evaporation at 
$B\simeq330\,$G, where the interstate scattering length is $-280\,a_0$, we are able to enter the Fermi degenerate regime 
($T_{\rm Li}/T_F \lesssim 0.6$) with total number $N_{\rm Li}=1.5 \times 10^5$. Applying this approach to the Li-Yb 
mixture however leads to reduced initial $N_{\rm Yb}$ (see Fig. 1 (b)) and shorter Yb lifetime from the sympathetic 
evaporation effect. We therefore restrict ourselves to keeping the initial ratio $N_{\rm Yb}/N_{\rm Li}$ large.

We now describe our measurements of sympathetic cooling of \li in contact with \yb which is undergoing forced evaporative 
cooling through a continuous lowering of the trap depth. In order to boost the collision rate, the single beam ODT for the 
thermalization measurement is modified to a crossed beam geometry by adding a second laser beam which intersects the first 
at a shallow angle of about $10^{\circ}$, and has the same power, orthogonal polarization, and a larger waist of $50\mu$m. 
After loading the atoms, we reduce the power in the optical trap following an approximately exponential shape. 
Fig.\ref{fig:symp} shows the number and temperature evolution during such an evaporation ramp at near-zero magnetic 
field.

We observe that Yb decays quickly with an exponential time constant of $4.3\,$s, while the longer decay constant of 
$24\,$s for Li is comparable with the vacuum limited lifetime. This implies that Yb loss is primarily from 
trap depth reduction while Li loss is primarily from background processes, as desired for efficient sympathetic cooling. 
During the cooling process, the Li phase space density increases by about three orders of magnitude as the gas is brought 
below the Fermi temperature, to $T_{\rm Li}=1.2\,\mu$K with $T_{\rm Li}/T_{F}\simeq 0.7$. At this point 
$T_{\rm Yb}=650\,$nK, a factor of four above the critical temperature for Bose condensation. We are prevented from further 
sympathetic cooling by the rapidly increasing inter-species thermalization time from the lowered densities and lowered 
overlap from unequal gravitational sag. We are currently improving our cooling scheme by implementing more tightly focused 
ODT beams. Substantial improvements are also expected from the introduction of a magnetic field gradient to allow 
manipulation of the Li trap depth and position, without affecting Yb.  

Our results establish a stable ultracold alkali+spin singlet mixture and also constitute the first instance of sympathetic 
cooling of a second atomic species by a spin-singlet atom. Future work includes studies of molecular levels by one- and 
two-photon photoassociation spectroscopies and searches for magnetically and optically induced Feshbach resonances 
\cite{zuch10,ciur05}, important steps towards production of paramagnetic polar molecules of LiYb. Improving our 
sympathetic cooling arrangement with magnetic gradients and tighter beams will allow us to reach double
quantum degeneracy with various combinations of Li and Yb isotopes, including Fermi-Fermi degenerate mixtures with high
mass imbalance. Yb atoms inside a degenerate Li cloud can also serve as an impurity to study superfluidity
\cite{impuritynote}, and for thermometry of a deeply degenerate Fermi gas \cite{spie09}. 

We thank W. Willcockson, J.K. Smith, R. Weh, and W. English for major technical contributions during the 
initial stages of the experiment, and A. G$\ddot{\rm o}$rlitz and D.M. Stamper-Kurn for helpful comments. We gratefully 
acknowledge support from the National Science Foundation, Sloan Foundation, UW Royalty Research Fund, and NIST. A.K. 
thanks the NSERC. F.M. thanks the DAAD.


\begin{thebibliography}{31}
\bibitem[{MOT(2010)}]{MOTinelnote}
\bibinfo{journal}{We observe significant reduction of Yb MOT number when
  operating simultaneous Li and Yb MOTs. The optimum MOT gradient for each
  species is related to the natural linewidth of the cooling transition and for
  our specific conditions differs empirically by a factor of 7.}
\bibitem[{imp(2010)}]{impuritynote}
\bibinfo{journal}{While the Li cloud is smaller than an equal temperature Yb
  cloud in the Boltzmann regime, because of Fermi pressure the opposite is true
  for $T/T_F \lesssim 0.27$ in the unitary regime of strong \li interactions.}
\end{thebibliography}
\end{document}